\documentclass[10pt]{article}
\usepackage{amsmath}
\usepackage{amssymb}
\usepackage{graphicx}
\usepackage{cite}
\usepackage{color}
\usepackage[labelfont=bf,labelsep=period,justification=raggedright]{caption}

\makeatletter
\renewcommand{\@biblabel}[1]{\quad#1.}
\makeatother
\newcommand{\be}{\begin{equation}}
\newcommand{\ee}{\end{equation}}
\newcommand{\bea}{\begin{eqnarray}}
\newcommand{\eea}{\end{eqnarray}}

\input{tcilatex}
\begin{document}

\date{}


\begin{flushleft}
{\Large \textbf{Analysis of two-player quantum games in an EPR setting} } 
\newline
James M.~Chappell$^{1,2,\ast}$, Azhar Iqbal$^{2}$, Derek Abbott$^{2}$ 
\newline
\textbf{1} School of Chemistry and Physics, University of Adelaide, South Australia,
Australia \newline
\textbf{2} School of Electrical and Electronic Engineering, University of
Adelaide, South Australia, Australia \newline
$\ast$ E-mail: james.m.chappell@adelaide.edu.au
\end{flushleft}


\section*{Abstract}

The framework for playing quantum games in an Einstein-Podolsky-Rosen (EPR)
type setting is investigated using the mathematical formalism of Clifford
geometric algebra (GA). In this setting, the players' strategy sets remain
identical to the ones in the classical mixed-strategy version of the game,
which is then obtained as proper subset of the corresponding quantum game.
As examples, we analyze the games of Prisoners' Dilemma and Stag
Hunt when played in the EPR type setting.


\section*{Introduction}

Although its origins can be traced to earlier works \cite%
{Blaquiere,Wiesner,Mermin,Mermin1}, the extension of game theory \cite%
{Binmore,Rasmusen} to the quantum regime \cite{Peres} was proposed by Meyer 
\cite{MeyerDavid} and Eisert et al \cite{Eisert1999} and has since been
investigated by others \cite%
{Vaidman,BenjaminHayden,EnkPike2002,Johnson,MarinattoWeber,IqbalToor1,DuLi,Du,Piotrowski,FlitneyAbbottRoyal,FlitneyAbbott1,IqbalToor2,Piotrowski1,Shimamura1,FlitneyAbbott2,IqbalWeigert,Mendes,CheonTsutsui,IqbalEpr:2005,NawazToor,CheonAIP,Shimamura,IchikawaTsutsui,OzdemirShimamura,IqbalCheon,Ramzan,FlitneyHollenberg2006,Aharon,Bleiler,ahmed2008three,GuoZhang,Ichikawa,IqbalCheonAbbott,Qiang,IqbalAbbott,iqbal2009quantum,CIL,NFJP2010,Chappell3Player}%
. Game theory is a vast subject but many interesting strategic interactions
can still be found in simple-to-analyze two-player two-strategy
non-cooperative games. The well known games of Prisoners' Dilemma (PD) and
Stag Hunt \cite{Binmore,Rasmusen} are two such examples.

The general idea in the quantization scheme proposed by Eisert et al \cite%
{Eisert1999} for such games involves a referee who forwards a two-qubit entangled
state to the two players. Players perform their strategic actions on the
state that consist of local unitary transformations to their respective
qubits. The qubits are subsequently returned to the referee for measurement
from which the players' payoffs are determined. The setup ensures that
players sharing a product initial state corresponds to the mixed-strategy
version of the considered classical game. However, players sharing an
entangled state can lead to new Nash equilibria (NE) \cite{Binmore,Rasmusen}
consisting of pairs of unitary transformations \cite{Peres,Eisert1999}. At these
quantum NE the players can have higher payoffs relative to what they obtain
at the NE in the mixed-strategy version of the classical game.

This approach to constructing quantum games was subsequently criticized \cite%
{EnkPike2002} as follows. The players' strategic actions in the quantum game are
extended operations relative to their actions in the original mixed-strategy
version of the classical game, in which, each player can perform a strategic
action consisting of a probabilistic combination of their two pure
strategies. The mentioned criticism \cite{EnkPike2002} argued that as the
quantum players have expanded strategy sets and can do more than what the
classical players can do, it is plausible to represent the quantum game as
an extended classical game that also involves new pure strategies. The
entries in the extended game matrix can then be suitably chosen so to be
representative of the players' payoffs at the obtained quantum NE. This line
of reasoning can be extended further in stating that quantum games are in
fact `disguised' classical games and to quantize a game is equivalent to
replacing the original game by an extended classical game.

As a way to counter the criticism in \cite{EnkPike2002}, two-party
Einstein-Podolsky-Rosen (EPR) type experiments \cite%
{EPR,Bohm,Bell,Bell1,Bell2,Aspect,ClauserShimony,Cereceda} are recognized to
have genuinely quantum features. One observes that the setting of such
experiments can be fruitfully adapted \cite%
{IqbalWeigert,IqbalEpr:2005,IqbalCheon,IqbalCheonAbbott,iqbal2009quantum} for playing
a quantum version of a two-player two-strategy game, which allows us to
avoid the criticism from another perspective. In particular, with the EPR
type setting the players' strategies can be defined entirely
classically---consisting of a probabilistic combination of a player's choice
between two measurement directions. That is, with this setting, the players'
strategy sets remain identical to ones they have in a standard arrangement
for playing a mixed-strategy version of a classical two-player two-strategy
game. As the players' strategy sets in the quantum game are not extended
relative to the classical game, for this route to constructing quantum
games, the mentioned criticism \cite{EnkPike2002} does not apply.

The usefulness of applying the formalism of geometric algebra (GA) \cite%
{GA1,GA,Doran:2003,Venzo2007,DoranParker:2001,chappell2011geometric,chappell2011revisiting} in the
investigation of quantum games has recently been shown \cite{CIL} for
the well known quantum penny flip game \cite{MeyerDavid}. One may ask about
the need of using the formalism of GA when, for instance, the GA based
analysis of two-player quantum games developed in the following can also be
reproduced with the standard analysis with Pauli matrices. We argue that the
Pauli matrices are not always the preferred representation. Especially, as
it is quite often overlooked that the algebra of Pauli matrices is the
matrix representation for the Clifford's geometric algebra $\mathcal{R}^{3}$%
, which is no more and no less than a system of directed numbers
representing the geometrical properties of Euclidean $3$-space. As a GA
based analysis allows using operations in $3$-space with \textit{real
coordinates}, it thus permits a visualization that is simply not available
in the standard approach using matrices over the field of complex numbers.
Pauli matrices are isomorphic to the quaternions, and hence represent
rotations of particle states. This fact paves the way to describe general
unitary transformations on qubits, in a simplified algebraic form, as 
\textit{rotors} that bring noticeable simplifications and geometrical
clarifications. We apply constraints on the parameters of EPR type
arrangements that ensure a faithful embedding of the mixed-strategy version
of the original classical game within the corresponding quantum game. In
particular, we show how using GA we can determine new NE in quantum games of
Stag Hunt and Prisoners' Dilemma played in the EPR type setting.

\subsection*{EPR setting for playing a quantum game}

We have the following payoff matrices 
\begin{equation}
\mathcal{A}=%
\begin{array}{c}
\text{Alice}%
\end{array}%
\begin{array}{c}
S_{1} \\ 
S_{2}%
\end{array}%
\overset{\overset{%
\begin{array}{c}
\text{Bob}%
\end{array}%
}{%
\begin{array}{cc}
S_{1}^{\prime } & S_{2}^{\prime }%
\end{array}%
}}{\left( 
\begin{array}{cc}
G_{00} & G_{01} \\ 
G_{10} & G_{11}%
\end{array}%
\right) },\text{ \ \ }\mathcal{B}=%
\begin{array}{c}
\text{Alice}%
\end{array}%
\begin{array}{c}
S_{1} \\ 
S_{2}%
\end{array}%
\overset{\overset{%
\begin{array}{c}
\text{Bob}%
\end{array}%
}{%
\begin{array}{cc}
S_{1}^{\prime } & S_{2}^{\prime }%
\end{array}%
}}{\left( 
\begin{array}{cc}
H_{00} & H_{01} \\ 
H_{10} & H_{11}%
\end{array}%
\right) },  \label{AandBmatrices}
\end{equation}%
giving Alice's and Bob's payoffs, respectively. Here Alice's pure strategies
are $S_{1}$ and $S_{2}$ and Bob's pure strategies are $S_{1}^{\prime }$ and $%
S_{2}^{\prime }$. In a run, Alice chooses her strategy to be either $S_{1}$
or $S_{2}$ and likewise, in the same run, Bob chooses his strategy to be
either $S_{1}^{\prime }$ or $S_{2}^{\prime }$. We consider games with
symmetrical payoffs for which $\mathcal{B}=\mathcal{A}^{T}$, where $T$
indicates transpose. This requires $H_{00}=G_{00},$ $H_{01}=G_{10},$ $%
H_{10}=G_{01},$ and $H_{11}=G_{11}.$

The EPR setting assumes that players Alice and Bob are spatially-separated
participants, who are located at the two arms of the EPR system. In a run,
each player receives one half of a two-particle system emitted by the same
source. We associate Alice's strategies $S_{1},S_{2}$ to the directions $%
\kappa _{1}^{1},\kappa _{2}^{1}$ respectively and similarly, associate Bob's
strategies $S_{1}^{\prime },S_{2}^{\prime }$ to the directions $\kappa
_{1}^{2},\kappa _{2}^{2}$, respectively. On receiving a pair of particles,
players Alice and Bob together choose a pair of directions from the four
possible cases $(\kappa _{1}^{1},\kappa _{1}^{2}),$ $(\kappa _{1}^{1},\kappa
_{2}^{2}),$ $(\kappa _{2}^{1},\kappa _{1}^{2}),$ $(\kappa _{2}^{1},\kappa
_{2}^{2})$ and a quantum measurement is performed along the chosen pair. The
outcome of the measurement at either arm is $+1$ or $-1$. Over a large
number of runs, a record is maintained of the players' choices of
directions, representing their strategies, and one of the four possible
outcomes $(+1,+1),$ $(+1,-1),$ $(-1,+1),$ $(-1,-1)$ emerging out of the
measurement. Within each of the brackets, the first entry is reserved for
the outcome at Alice's side and the second entry for the outcome at Bob's
side. Players' payoff relations are expressed in terms of the outcomes of
measurements that are recorded for a large number of runs, as the players
sequentially receive, two-particle systems emitted from the source. These
payoffs depend on the strategic choices that each player adapts for his/her
two directions over many runs, and on the dichotomic outcomes of the
measurements performed along those directions.

\subsection*{Geometric algebra}

Geometric algebra (GA) \cite{GA1,GA,Doran:2003,Venzo2007,DoranParker:2001} is an associative
non-commutative algebra, that can provide an equivalent description to the
conventional Dirac bra-ket and matrix formalisms of quantum mechanics,
consisting of solely of algebraic elements over a strictly real field.
Recently, Christian \cite{Christian,christian2011restoring} has used the formalism of GA in thought
provoking investigations of some of the foundational questions in quantum
mechanics. In the area of quantum games, GA has been used by Chappell et al 
\cite{CIL} to determine all possible unitary transformations that
implement a winning strategy in Meyer's PQ penny flip quantum game \cite%
{MeyerDavid}, and also in analyzing three-player quantum games \cite%
{Chappell3Player}.

Given a linear vector space $V$ with elements $u,v,\dots $ we may form \cite%
{Szekeres:2004} the tensor product $U\otimes V$ of vector spaces $U,V$,
containing elements (bivectors) $u\otimes v$ and hence construct the
exterior or wedge product $u\wedge v=u\otimes v-v\otimes u$. This may be
extended to a vector space $\Lambda (V)$ with elements consisting of
multivectors that can be multiplied by means of the exterior product. The
geometric product $uv$ of two vectors $u,v$ is defined by $uv=u.v+u\wedge v$%
, where $u.v$ is the scalar inner product. The geometric product is in
general not commutative but it is always associative, that is $u(vw)=(uv)w$.

We denote by $\{\sigma _{i}\}$ an orthonormal basis in $\Re ^{3}$, then $%
\sigma _{i}\cdot \sigma _{j}=\delta _{ij}$. We also have $\sigma _{i}\wedge
\sigma _{i}=0$ for each $i=1,2,3$ and so in terms of the geometric product
we have $\sigma _{i}^{2}=\sigma _{i}\sigma _{i}=1$, and $\sigma _{i}\sigma
_{j}=\sigma _{i}\wedge \sigma _{j}=-\sigma _{j}\sigma _{i}$ for each $i\neq
j $. Hence the basis vectors anticommute with respect to the geometric
product. If we denote by $\iota $ the trivector 
\begin{equation}
\iota =\sigma _{1}\sigma _{2}\sigma _{3},
\end{equation}%
then for distinct basis vectors we have 
\begin{equation}
\sigma _{i}\sigma _{j}=\delta _{ij}+\iota \epsilon _{ijk}\sigma _{k},
\end{equation}%
where $\epsilon _{ijk}$ is the Levi-Civita symbol. We find that $\iota
^{2}=\sigma _{1}\sigma _{2}\sigma _{3}\sigma _{1}\sigma _{2}\sigma
_{3}=\sigma _{1}\sigma _{2}\sigma _{1}\sigma _{2}=-1$ and commutes with all
other elements and so has identical properties to the conventional complex
number $\mathrm{i}=\sqrt{-1}$. Thus we have an isomorphism between the basis
vectors $\sigma _{1},\sigma _{2},\sigma _{3}$ and the Pauli matrices through
the use of the geometric product.

In order to express quantum states in GA we use the one-to-one mapping \cite%
{Doran:2003,Venzo2007,DoranParker:2001} defined as follows 
\begin{equation}
|\psi \rangle =\alpha |0\rangle +\beta |1\rangle =%
\begin{bmatrix}
a_{0}+\mathrm{i}a_{3} \\ 
-a_{2}+\mathrm{i}a_{1}%
\end{bmatrix}%
\leftrightarrow \psi =a_{0}+a_{1}\iota \sigma _{1}+a_{2}\iota \sigma
_{2}+a_{3}\iota \sigma _{3},  \label{eq:spinorMapping}
\end{equation}%
where $a_{i}$ are real scalars.

It can then be shown using the Schmidt decomposition of a general two qubit
state \cite{DoranParker:2001}, that a general two-particle state can be represented 
in GA as 
\begin{equation}
\psi =AB(\cos \frac{\gamma }{2}+\sin \frac{\gamma }{2}{\iota }\sigma _{2}^{1}%
{\iota }\sigma _{2}^{2}),  \label{eq:genEntM1}
\end{equation}%
where $\gamma \in \lbrack 0,\frac{\pi }{2}]$ is a measure of the
entanglement and where $A,B$ are single particle rotors applied to the first
and second qubit, respectively. General unitary operations are called \cite%
{Doran:2003} rotors in GA, represented as 
\begin{equation}
\mathbf{R}(\theta _{1},\theta _{2},\theta _{3})=\mathrm{e}^{-\theta _{3}{%
\iota }\sigma _{3}/2}\mathrm{e}^{-\theta _{1}{\iota }\sigma _{2}/2}\mathrm{e}%
^{-\theta _{2}{\iota }\sigma _{3}/2}.  \label{eq:GenUnitaryRotation}
\end{equation}%
This rotation, in Euler angle form, can completely explore the available
space of a single qubit, and is equivalent to a general unitary
transformation acting on a spinor. So, we have the rotors for each qubit
defined as 
\begin{eqnarray}
A &=&\mathbf{R}(\alpha _{1},\alpha _{2},\alpha _{3})=\mathrm{e}^{-\alpha _{3}%
{\iota }\sigma _{3}/2}\mathrm{e}^{-\alpha _{1}{\iota }\sigma _{2}/2}\mathrm{e%
}^{-\alpha _{2}{\iota }\sigma _{3}/2}, \\
B &=&\mathbf{R}(\beta _{1},\beta _{2},\beta _{3})=\mathrm{e}^{-\beta _{3}{%
\iota }\sigma _{3}/2}\mathrm{e}^{-\beta _{1}{\iota }\sigma _{2}/2}\mathrm{e}%
^{-\beta _{2}{\iota }\sigma _{3}/2}.
\end{eqnarray}%
For example, for $A=B=1$ and $\gamma =\frac{\pi }{2}$, we find the Bell
state, and $A=1$ and $B=\mathbf{R}(\pi ,0,0)$ and $\gamma =\frac{\pi }{2}$
we recover the singlet state. This can be checked using Eq.~(\ref%
{eq:spinorMapping}), where we note that $-\iota \sigma _{2}\rightarrow
|1\rangle $.

To simulate the process of measurement in GA, we form a separable state $%
\phi =RS$, where $R$ and $S$ are single particle rotors, which allow general
measurement directions to be specified, on the first and second qubit
respectively. The state to be measured is now projected onto the separable
state $\phi $. In the $N$-particle case, the probability that the quantum
state $\psi $ returns the separable state $\phi $ is given is Ref. \cite%
{Doran:2003} as 
\begin{equation}
P(\psi ,\phi )=2^{N-2}\left( \langle \psi E\psi ^{\dagger }\phi E\phi
^{\dagger }\rangle _{0}-\langle \psi J\psi ^{\dagger }\phi J\phi ^{\dagger
}\rangle _{0}\right) ,  \label{eq:DoranOverlapProb}
\end{equation}%
where the angle brackets $\left\langle \cdot \right\rangle _{0}$ mean to
retain only the scalar part of the expression. We have the two observables $%
\psi J\psi ^{\dagger }$ and $\psi E\psi ^{\dagger }$, which in the two
particle case involves \cite{Doran:2003} 
\begin{equation}
E=\frac{1}{2}(1-{\iota }\sigma _{3}^{1}{\iota }\sigma _{3}^{2}),\quad J=%
\frac{1}{2}({\iota }\sigma _{3}^{1}+{\iota }\sigma _{3}^{2}).
\label{eq:DoranOverlapE}
\end{equation}%
The $\dagger $ operator acts in the same way as complex conjugation,
flipping the sign of $\iota $ and inverting the order of terms.

\section*{Results}

Employing Eq.~(\ref{eq:DoranOverlapProb}), we firstly calculate 
\begin{align}
\psi E\psi ^{\dagger }& =\frac{1}{2}AB(\cos \frac{\gamma }{2}+\sin \frac{%
\gamma }{2}{\iota }\sigma _{2}^{1}{\iota }\sigma _{2}^{2})(1-{\iota }\sigma
_{3}^{1}{\iota }\sigma _{3}^{2})(\cos \frac{\gamma }{2}+\sin \frac{\gamma }{2%
}{\iota }\sigma _{2}^{1}{\iota }\sigma _{2}^{2})B^{\dagger }A^{\dagger } 
\notag \\
& =\frac{1}{2}AB\left( 1-{\iota }\sigma _{3}^{1}{\iota }\sigma _{3}^{2}+\sin
\gamma ({\iota }\sigma _{2}^{1}{\iota }\sigma _{2}^{2}-{\iota }\sigma
_{1}^{1}{\iota }\sigma _{1}^{2})\right) B^{\dagger }A^{\dagger }  \notag \\
& =\frac{1}{2}\left( 1-{\iota }A\sigma _{3}^{1}A^{\dagger }{\iota }B\sigma
_{3}^{2}B^{\dagger }+\sin \gamma ({\iota }A\sigma _{2}^{1}A^{\dagger }{\iota 
}B\sigma _{2}^{2}B^{\dagger }-{\iota }A\sigma _{1}^{1}A^{\dagger }{\iota }%
B\sigma _{1}^{2}B^{\dagger })\right)  \label{eq:GeneralEObservable2}
\end{align}%
and 
\begin{align}
\psi J\psi ^{\dagger }& =\frac{1}{2}AB(\cos \frac{\gamma }{2}+\sin \frac{%
\gamma }{2}{\iota }\sigma _{2}^{1}{\iota }\sigma _{2}^{2})({\iota }\sigma
_{3}^{1}+{\iota }\sigma _{3}^{2})(\cos \frac{\gamma }{2}+\sin \frac{\gamma }{%
2}{\iota }\sigma _{2}^{1}{\iota }\sigma _{2}^{2})B^{\dagger }A^{\dagger } 
\notag \\
& =\frac{1}{2}AB(\cos ^{2}\frac{\gamma }{2}-\sin ^{2}\frac{\gamma }{2})({%
\iota }\sigma _{3}^{1}+{\iota }\sigma _{3}^{2})B^{\dagger }A^{\dagger } 
\notag \\
& =\frac{1}{2}\cos \gamma ({\iota }A\sigma _{3}^{1}A^{\dagger }+{\iota }%
B\sigma _{3}^{2}B^{\dagger }).  \label{eq:GeneralJObservable2}
\end{align}%
To describe the players measurement directions, we have $R=e^{-{\iota }%
\kappa ^{1}\sigma _{2}^{1}}$ and $S=e^{-{\iota }\kappa ^{2}\sigma _{2}^{2}}$%
. For the quantum game in the EPR setting, $\kappa ^{1}$ can be either of
Alice's two directions i.e. $\kappa _{1}^{1}$ or $\kappa _{2}^{1}$.
Similarly, in the expression for $S$ the $\kappa ^{2}$ can be either of
Bob's two directions i.e. $\kappa _{1}^{2}$ or $\kappa _{2}^{2}$.\emph{\ }%
Hence we obtain 
\begin{align}
\phi J\phi ^{\dagger }& =RSJS^{\dagger }R^{\dagger }  \notag \\
& =\frac{1}{2} \left ({\iota }R\sigma _{3}^{1}R^{\dagger }+{\iota }S\sigma
_{3}^{2}S^{\dagger } \right )  \notag \\
& =\frac{1}{2} \left ({\iota }\sigma _{3}^{1}e^{{\iota }\kappa ^{1}\sigma _{2}^{1}}+%
{\iota }\sigma _{3}^{2}e^{{\iota }\kappa ^{2}\sigma _{2}^{2}} \right ),
\end{align}%
and 
\begin{align}
\phi E\phi ^{\dagger }& =RSES^{\dagger }R^{\dagger }  \notag \\
& =\frac{1}{2} \left (1-{\iota }R\sigma _{3}^{1}R^{\dagger }{\iota }S\sigma
_{3}^{2}S^{\dagger } \right )  \notag \\
& =\frac{1}{2} \left (1-{\iota }\sigma _{3}^{1}e^{{\iota }\kappa ^{1}\sigma
_{2}^{1}}{\iota }\sigma _{3}^{2}e^{{\iota }\kappa ^{2}\sigma _{2}^{2}} \right ).
\end{align}%
Now from Eq.~(\ref{eq:DoranOverlapProb}), we calculate 
\begin{align}
-\left \langle \psi J\psi ^{\dagger }\phi J\phi ^{\dagger } \right \rangle _{0}& =-\frac{1%
}{4} \left \langle \cos \gamma \left ({\iota }A\sigma _{3}^{1}A^{\dagger }+{\iota }%
B\sigma _{3}^{2}B^{\dagger } \right ) \left ({\iota }\sigma _{3}^{1}e^{{\iota }\kappa
^{1}\sigma _{2}^{1}}+{\iota }\sigma _{3}^{2}e^{{\iota }\kappa ^{2}\sigma
_{2}^{2}} \right ) \right \rangle _{0}  \notag \\
& =\frac{1}{4}\cos \gamma \left [ (-)^{m}X(\kappa ^{1})+(-)^{n}Y(\kappa^{2}) \right ],  \label{eq:Method1MeasureJBasic}
\end{align}%
where $m,n\in \{0,1\}$ refers to measuring a $|0\rangle $ or a $|1\rangle $
state, respectively, and using the results in Appendix, we have 
\begin{eqnarray}
X(\kappa ^{1}) &=&\cos \alpha _{1}\cos \kappa ^{1}+\cos \alpha _{3}\sin
\alpha _{1}\sin \kappa ^{1}, \\
Y(\kappa ^{2}) &=&\cos \beta _{1}\cos \kappa ^{2}+\cos \beta _{3}\sin \beta
_{1}\sin \kappa ^{2}.  \label{eq:XandYRelations}
\end{eqnarray}%
Also, from Eq.~(\ref{eq:DoranOverlapProb}) we obtain 
\begin{align}
& \left  \langle \psi E\psi ^{\dagger }\phi E\phi ^{\dagger } \right \rangle _{0}  \\
& \;\;\; \;\;\;= \Big \langle
(1-{\iota }A\sigma _{3}^{1}A^{\dagger }{\iota }B\sigma _{3}^{2}B^{\dagger
}+\sin \gamma ({\iota }A\sigma _{2}^{1}A^{\dagger }{\iota }B\sigma
_{2}^{2}B^{\dagger }-{\iota }A\sigma _{1}^{1}A^{\dagger }{\iota }B\sigma
_{1}^{2}B^{\dagger }))  \notag \\
& \;\;\; \;\;\; \times (1-{\iota }\sigma _{3}^{1}{\iota }\sigma _{3}^{2}e^{{\iota }\kappa
\sigma _{2}^{1}}e^{{\iota }\tau \sigma _{2}^{2}}) \Big \rangle _{0}  \notag \\
& \;\;\; \;\;\; =\frac{1}{4} \left [1+(-)^{m+n}XY-(-)^{m+n}\sin \gamma
\{U(k^{1})V(k^{2})-F(k^{1})G(k^{2})\} \right ],  \label{eq:Method1MeasureEBasic}
\end{align}
where 
\begin{align}
F(\kappa ^{1})& =\cos \alpha _{2}(\cos \kappa ^{1}\sin \alpha _{1}-\cos
\alpha _{3}\sin \kappa ^{1}\cos \alpha _{1})+\sin \kappa ^{1}\sin \alpha
_{2}\sin \alpha _{3}, \\
G(\kappa ^{2})& =\cos \beta _{2}(\cos \kappa ^{2}\sin \beta _{1}-\cos \beta
_{3}\sin \kappa ^{2}\cos \beta _{1})+\sin \kappa ^{2}\sin \beta _{2}\sin
\beta _{3}  \label{eq:FGHRelations1}
\end{align}%
and 
\begin{align}
U(\kappa ^{1})& =-\sin \alpha _{2}(\cos \kappa ^{1}\sin \alpha _{1}-\cos
\alpha _{3}\sin \kappa ^{1}\cos \alpha _{1})+\sin \kappa ^{1}\cos \alpha
_{2}\sin \alpha _{3}, \\
V(\kappa ^{2})& =-\sin \beta _{2}(\cos \kappa ^{2}\sin \beta _{1}-\cos \beta
_{3}\sin \kappa ^{2}\cos \beta _{1})+\sin \kappa ^{2}\cos \beta _{2}\sin
\beta _{3}.  \label{eq:FGHRelations2}
\end{align}%
To simplify the equations, we define 
\begin{align}
Z(\kappa ^{1},\kappa ^{2})& =F(k^{1})G(k^{2})-U(k^{1})V(k^{2})
\label{eq:ZEquation-1} \\
& =\cos \phi \lbrack \cos \kappa ^{1}\cos \kappa ^{2}\sin \beta _{1}\sin
\alpha _{1}-\sin \kappa ^{1}\cos \kappa ^{2}\sin \beta _{1}\cos \alpha
_{1}\cos \alpha _{3}  \notag \\
& +\sin \kappa ^{1}\sin \kappa ^{2}(\cos \alpha _{1}\cos \alpha _{3}\cos
\beta _{1}\cos \beta _{3}-\sin \alpha _{3}\sin \beta _{3})  \notag \\
& -\cos \kappa ^{1}\sin \kappa ^{2}\sin \alpha _{1}\cos \beta _{1}\cos \beta
_{3}]  \notag \\
& +\sin \phi \lbrack \sin \kappa ^{1}\cos \kappa ^{2}\sin \alpha _{3}\sin
\beta _{1}+\cos \kappa ^{1}\sin \kappa ^{2}\sin \alpha _{1}\sin \beta _{3} 
\notag \\
& +\sin \kappa ^{1}\sin \kappa ^{2}(\cos \beta _{1}\cos \beta _{3}\sin
\alpha _{3}+\cos \alpha _{1}\cos \alpha _{3}\sin \beta _{3})].
\label{eq:ZEquation}
\end{align}%
Now combining Eq.~(\ref{eq:Method1MeasureJBasic}) and Eq.~(\ref%
{eq:Method1MeasureEBasic}) we have the probability to observe a particular
state 
\begin{equation}
P_{mn}=\frac{1}{4} \left [1+\cos \gamma
\{(-)^{m}X_{i}+(-)^{n}Y_{j}\}+(-)^{m+n}(X_{i}Y_{j}+\sin \gamma Z_{ij}) \right ].
\label{eq:Method1Prob}
\end{equation}%
To simplify notation we have written $Z_{ij}=Z(\kappa _{i}^{1},\kappa
_{j}^{2})$ , $X_{i}=X(\kappa _{i}^{1})$ and $Y_{j}=Y(\kappa _{j}^{2})$,
where $i,j\in \{1,2\}$ represent the two possible measurement directions
available to each player. If we put $\gamma =0$, that is, for no entanglement, we have the
probability 
\begin{align}
P_{mn}& =\frac{1}{4} \left (1+(-)^{m}X_{i}+(-)^{n}Y_{j}+(-)^{m+n}X_{i}Y_{j} \right )  \notag
\\
& =\frac{(1+(-)^{m}X_{i})^{1}}{2}\frac{(1+(-)^{n}Y_{j})^{2}}{2},
\end{align}%
which shows a product state incorporating general measurement directions for
each qubit.

Writing out the probabilities for the four measurement outcomes we find 
\begin{align}
P_{00}(\kappa _{i}^{1},\kappa _{j}^{2})& =\frac{1}{4} \left [1+\cos \gamma
(X_{i}+Y_{j})+(X_{i}Y_{j}+\sin \gamma Z_{ij}) \right ],  \label{probability00} \\
P_{01}(\kappa _{i}^{1},\kappa _{j}^{2})& =\frac{1}{4}\left [1+\cos \gamma
(X_{i}-Y_{j})-(X_{i}Y_{j}+\sin \gamma Z_{ij}) \right ],  \label{probability01} \\
P_{10}(\kappa _{i}^{1},\kappa _{j}^{2})& =\frac{1}{4} \left [1+\cos \gamma
(-X_{i}+Y_{j})-(X_{i}Y_{j}+\sin \gamma Z_{ij}) \right ],  \label{probability10} \\
P_{11}(\kappa _{i}^{1},\kappa _{j}^{2})& =\frac{1}{4} \left [1+\cos \gamma
(-X_{i}-Y_{j})+(X_{i}Y_{j}+\sin \gamma Z_{ij}) \right ].  \label{probability11}
\end{align}

\subsection*{Finding the payoff relations}

We allow each player the classical probabilistic choice between their two
chosen measurement directions for their Stern-Gerlach detectors. The two
players, Alice and Bob choose their first measurement direction with
probability $x$ and $y$ respectively, where $x,y\in \lbrack 0,1]$. Now, we
have the mathematical expectation of Alice's payoff, where she chooses the
direction $\kappa _{1}^{1}$ with probability $x$ and the measurement
direction $\kappa _{2}^{1}$ with probability $1-x$, as 
\begin{align}
\Pi _{A}(x,y) & =xy[P_{00} G_{00}+P_{01} G_{01}+P_{10} G_{10}+P_{11} G_{11}]  \notag \\
& +x(1-y)[P_{00} G_{00}+P_{01} G_{01}+P_{10} G_{10}+P_{11} G_{11}]  \notag \\
& +y(1-x)[P_{00} G_{00}+P_{01} G_{01}+P_{10} G_{10}+P_{11} G_{11}]  \notag \\
& +(1-x)(1-y)[P_{00} G_{00}+P_{01} G_{01}+P_{10} G_{10}+P_{11} G_{11}],  \label{EPRpayoffsAlice}
\end{align}%
where we have used the payoff matrix, defined for Alice, in Eq.~(\ref%
{AandBmatrices}) and the subscript $A$ refers to Alice. We also define

\begin{equation}
\Delta _{1}=G_{10}-G_{00},\text{ }\Delta _{2}=G_{11}-G_{01},\text{ }\Delta
_{3}=\Delta _{2}-\Delta _{1},  \label{deltas}
\end{equation}%
so that by using Eqs.~(\ref{probability00}-\ref{probability11}) the payoff
for Alice (\ref{EPRpayoffsAlice}) is expressed as 
\begin{align}
& \Pi _{A}(x,y) \notag \\
& =\frac{1}{4} \Big [G_{00}+G_{10}+G_{01}+G_{11}  \notag \\
& +\Delta _{3}\{x((X_{1}-X_{2})Y_{2}+(Z_{12}-Z_{22})\sin \gamma
)+y((Y_{1}-Y_{2})X_{2}+(Z_{21}-Z_{22})\sin \gamma )  \notag \\
& +xy\{(X_{1}-X_{2})(Y_{1}-Y_{2})+\sin \gamma
(Z_{11}+Z_{22}-Z_{12}-Z_{21})\}+X_{2}Y_{2}+Z_{22}\sin \gamma \}  \notag \\
& -\cos \gamma \{(\Delta _{1}+\Delta
_{2})((X_{1}-X_{2})x+X_{2})-\Delta_4 ((Y_{1}-Y_{2})y+Y_{2})\} \Big ],
\label{eq:AlicePayoffExpandedFourCoin}
\end{align}%
where $ \Delta_4 = G_{00}-G_{01}+G_{10}-G_{11} $.
Bob's payoff, when Alice plays $x$ and Bob plays $y$ can now be obtained by
interchanging $x$ and $y$ in the right hand side of Eq.~(\ref%
{eq:AlicePayoffExpandedFourCoin}).

\subsection*{Solving the general two-player game}

We now find the optimal solutions by calculating the Nash equilibrium (NE),
that is, the expected response assuming rational self interest. To find the
NE we simply require 
\begin{equation}
\Pi _{A}(x^{\ast },y^{\ast })\geq \Pi _{A}(x,y^{\ast }),\quad \Pi
_{B}(x^{\ast },y^{\ast })\geq \Pi _{B}(x^{\ast },y),
\end{equation}%
which is stating that any unilateral movement of a player away from the NE
of $(x^{\ast },y^{\ast })$, will result in a lower payoff for that player.
We find 
\begin{align}
& \Pi _{A}(x^{\ast },y^{\ast })-\Pi _{A}(x,y^{\ast })  \notag \\
& =\frac{1}{4}(x^{\ast }-x) \Big [\Delta _{3} \big \{ y^{\ast
}((X_{1}-X_{2})(Y_{1}-Y_{2})+\sin \gamma (Z_{11}+Z_{22}-Z_{12}-Z_{21})) 
\notag \\
& +(X_{1}-X_{2})Y_{2}+(Z_{12}-Z_{22})\sin \gamma \big \}-\cos \gamma (\Delta
_{1}+\Delta _{2})(X_{1}-X_{2}) \Big ]  \label{eq:AliceGeneralNE}
\end{align}%
and for the second player Bob we have similarly 
\begin{align}
& \Pi _{B}(x^{\ast },y^{\ast })-\Pi _{B}(x^{\ast },y)  \notag \\
& =\frac{1}{4}(y^{\ast }-y) \Big [\Delta _{3} \big \{ x^{\ast
}((X_{1}-X_{2})(Y_{1}-Y_{2})+\sin \gamma (Z_{11}+Z_{22}-Z_{12}-Z_{21})) 
\notag \\
& +(Y_{1}-Y_{2})X_{2}+(Z_{21}-Z_{22})\sin \gamma \big \}-\cos \gamma (\Delta
_{1}+\Delta _{2})(Y_{1}-Y_{2}) \Big ].
\end{align}

\subsection*{Embedding the classical game}

To embed the classical game, we require at zero entanglement, not only the
same pair of strategies being a NE but also to have the bilinear structure
of the classical payoff relations. At a NE of $(x^{\ast },y^{\ast })=(0,0)$,
with zero entanglement, we find the payoff from Eq.~(\ref%
{eq:AlicePayoffExpandedFourCoin}) to be 
\begin{align}
\Pi _{A}(0,0)& =\frac{1}{4} \big [G_{00}(1+X_{2})(1+Y_{2})+G_{10}(1-X_{2})(1+Y_{2})
\notag \\
& +G_{01}(1+X_{2})(1-Y_{2})+G_{11}(1-X_{2})(1-Y_{2}) \big ].
\label{eq:AlicePayoffExpandedFourCoinZero1}
\end{align}%
This result illustrates how we could select any one of the payoff entries we
desire with the appropriate selection of $X_{2}$ and $Y_{2}$, however in
order to achieve the classical payoff of $G_{11}$ for this NE, we can see
that we require $X_{2}=-1$ and $Y_{2}=-1$. If we have a game which also has
a classical NE of $(x^{\ast },y^{\ast })=(1,1)$ then from Eq.~(\ref%
{eq:AlicePayoffExpandedFourCoin}) at zero entanglement we find the payoff 
\begin{align}
\Pi _{A}(1,1)& =\frac{1}{4} \big [G_{00}(1+X_{1})(1+Y_{1})+G_{10}(1-X_{1})(1+Y_{1})
\notag \\
& +G_{01}(1+X_{1})(1-Y_{1})+G_{11}(1-X_{1})(1-Y_{1}) \big ].
\label{eq:AlicePayoffExpandedFourCoinZero2}
\end{align}%
So, we can see, that we can select the required classical payoff, of $G_{00}$%
, by the selection of $X_{1}=1$ and $Y_{1}=1$.

Referring to Eq.~(\ref{eq:XandYRelations}), we then have the conditions 
\begin{eqnarray}
X(\kappa ^{1}) &=&\cos \alpha _{1}\cos \kappa ^{1}+\cos \alpha _{3}\sin
\alpha _{1}\sin \kappa ^{1}=\pm 1, \\
Y(\kappa ^{2}) &=&\cos \beta _{1}\cos \kappa ^{2}+\cos \beta _{3}\sin \beta
_{1}\sin \kappa ^{2}=\pm 1.
\end{eqnarray}%
Looking at the equation for Alice, we have two classes of solution: If $%
\alpha _{3}\neq 0$, then for the equations satisfying $X_{2}=Y_{2}=-1$, we
have for Alice in the first equation $\alpha _{1}=0$, $\kappa _{2}^{1}=\pi $
or $\alpha _{1}=\pi $, $\kappa _{2}^{1}=0$ and for the equations satisfying $%
X_{1}=Y_{1}=+1$, we have $\alpha _{1}=\kappa _{1}^{1}=0$ or $\alpha
_{1}=\kappa _{1}^{1}=\pi $, which can be combined to give either $\alpha
_{1}=0,$ $\kappa _{1}^{1}=0$ and $\kappa _{2}^{1}=\pi $ or $\alpha _{1}=\pi
, $ $\kappa _{1}^{1}=\pi $ and $\kappa _{2}^{1}=0$. For the second class
with $\alpha _{3}=0$, we have the solution $\alpha _{1}-\kappa _{2}^{1}=\pi $
and for $X_{1}=Y_{1}=+1$ we have $\alpha _{1}-\kappa _{1}^{1}=0$.

So, in summary, for both cases we have that the two measurement directions
are $\pi $ out of phase with each other, and for the first case ($\alpha
_{3}\neq 0$) we can freely vary $\alpha _{2}$ and $\alpha _{3}$, and for the
second case ($\alpha _{3}=0$), we can freely vary $\alpha _{1}$ and $\alpha
_{2}$ to change the initial quantum quantum state without affecting the game
NE or the payoffs. The same arguments hold for the equations for $Y$.
Combining these results and substituting into Eq.~(\ref{eq:ZEquation}), we
see by inspection for the two cases that

\begin{equation}
F(\kappa ^{1})=G(\kappa ^{2})=U(\kappa ^{1})=V(\kappa ^{2})=0,
\label{conditionsA}
\end{equation}%
and hence, we find that

\begin{equation}
Z_{22}=Z_{21}=Z_{12}=Z_{11}=0.  \label{consitionsB}
\end{equation}

This then reduces the equation governing the NE in Eq.~(\ref%
{eq:AliceGeneralNE}) to 
\begin{equation}
\Pi _{A}(x^{\ast },y^{\ast })-\Pi _{A}(x,y^{\ast })=\frac{1}{2}(x^{\ast
}-x) \big [\Delta _{3}\{2y^{\ast }-1\}-\cos \gamma (\Delta _{1}+\Delta _{2}) \big ]\geq
0,  \label{eq:NEEPREmbeddedStagHunt}
\end{equation}%
which now has the new quantum behavior governed solely by the entanglement
angle $\gamma $. We have the associated payoffs 
\begin{align}
\Pi _{A}(x,y)& =\frac{1}{2} \big [G_{00}+G_{11}-\cos \gamma
(G_{00}-G_{11})+2xy\Delta _{3}  \notag \\
& -x\{\Delta _{3}+\cos \gamma (\Delta _{1}+\Delta _{2})\}-y\{\Delta
_{3}-\cos \gamma (G_{00}-G_{01}+G_{10}-G_{11})\} \big ].
\label{eq:AlicePayoffQuantumGameClassical}
\end{align}%
Setting $\gamma =0$ in Eq.~(\ref{eq:AlicePayoffQuantumGameClassical}) we
find 
\begin{equation}
\Pi
_{A}(x,y)=G_{11}+x(G_{01}-G_{11})+y(G_{10}-G_{11})+xy(G_{00}-G_{01}-G_{10}+G_{11}),
\label{eq:AlicePayoffClassicalEmbedding}
\end{equation}%
which has the classical bilinear payoff structure in terms of $x$ and $y$.
Hence we have faithfully embedded the classical game inside a quantum
version of the game, when the entanglement goes to zero.

We also have the probabilities for each state $|m\rangle |n\rangle $, after
measurement from Eq.~(\ref{eq:Method1Prob}), for this form of the quantum
game as 
\begin{equation}
(P_{mn})_{ij}=\frac{1}{4} \left [1+\cos \gamma((-)^{m+i+1}+(-)^{n+j+1})+(-)^{m+n+i+j} \right ],
\label{eq:Method1ProbClassicalEmbedded}
\end{equation}%
for the two measurement directions $i$ and $j$.

\subsection*{Examples}

Here we explore the above results for the games of Prisoners' Dilemma and
Stag Hunt. The quantum versions of these games are discussed in Refs. \cite%
{Eisert1999,BenjaminHayden,FlitneyAbbottRoyal,FlitneyAbbott1,FlitneyAbbott2,IqbalAbbott}%
.

\subsubsection*{Prisoners' Dilemma}

The game of Prisoners' Dilemma (PD) \cite{Rasmusen} is widely known to
economists, social and political scientists and is one of the earliest games
to be investigated in the quantum regime \cite{Eisert1999}. PD describes the
following situation: two suspects are investigated for a crime that
authorities believe they have committed together. Each suspect is placed in
a separate cell and may choose between not confessing or confessing to have
committed the crime. Referring to the matrices (\ref{AandBmatrices}) we take 
$S_{1}\sim S_{1}^{\prime }$ and $S_{2}\sim S_{2}^{\prime }$ and identify $%
S_{1}$ and $S_{2}$ to represent the strategies of `not confessing' and
`confessing', respectively. If neither suspect confesses, i.e. $%
(S_{1},S_{1}) $, they go free, which is represented by $G_{00}$ units of
payoff for each suspect. The situation $(S_{1},S_{2})$ or $(S_{2},S_{1})$
represents in which one prisoner confesses while the other does not. In this
case, the prisoner who confesses gets $G_{10}$ units of payoff, which
represents freedom as well as financial reward as $G_{10}>G_{00}$, while the
prisoner who did not confess gets $G_{01}$, represented by his ending up in
the prison. When both prisoners confess, i.e. $(S_{2},S_{2})$, they both are
given a reduced term represented by $G_{11}$ units of payoff, where $%
G_{11}>G_{01}$, but it is not so good as going free i.e. $G_{00}>G_{11}$.

With reference to Eq.~(\ref{deltas}), we thus have $\Delta _{1},$ $\Delta
_{2}>0$. However, depending on the relative sizes of $\Delta _{1},$ $\Delta
_{2},$ the quantity $\Delta _{3}=\Delta _{2}-\Delta _{1}$ can be positive or
negative. At maximum entanglement ($\cos \gamma =0$), we note from Eq.~(\ref%
{eq:NEEPREmbeddedStagHunt}), that there are two cases depending on $\Delta
_{3}$. If $\Delta _{3}>0$, we notice that both the NE of $(x^{\ast },y^{\ast
})=(0,0)$ and $(x^{\ast },y^{\ast })=(1,1)$ are present, and from Eq.~(\ref%
{eq:AlicePayoffQuantumGameClassical}) we have the payoff in both cases 
\begin{equation}
\Pi _{A}(0,0)=\Pi _{B}(0,0)=\frac{1}{2}(G_{00}+G_{11})=\Pi _{A}(1,1)=\Pi
_{B}(1,1),
\end{equation}%
which is a significant improvement over the classical payoff of $G_{11}$.
For $\Delta _{3}<0$, we have the two NE of $(x^{\ast },y^{\ast })=(0,1)$ and 
$(x^{\ast },y^{\ast })=(1,0)$, and from Eq.~(\ref%
{eq:AlicePayoffQuantumGameClassical}) we have the payoff 
\begin{equation}
\Pi _{A}(0,1)=\Pi _{B}(0,1)=\frac{1}{2}(G_{01}+G_{10})=\Pi _{A}(1,0)=\Pi
_{B}(1,0).
\end{equation}%
If we reduce the entanglement of the qubits provided for the game,
increasing $\cos \gamma $ towards one, then from Eq.~(\ref%
{eq:NEEPREmbeddedStagHunt}), we find a phase phase transition to the
classical NE of $(x^{\ast },y^{\ast })=(0,0)$, at $\Delta _{3}-\cos \gamma
(\Delta _{1}+\Delta _{2})=0$ or 
\begin{equation}
\cos \gamma =\frac{\Delta _{3}}{\Delta _{1}+\Delta _{2}}=\frac{\Delta
_{2}-\Delta _{1}}{\Delta _{2}+\Delta _{1}}.
\end{equation}%
Because we know that $\Delta _{1},$ $\Delta _{2}>0$, for the PD game, then a
phase transition to the classical NE is guaranteed to occur, in the range $%
[0,1]$.

Consider a particular example of PD by taking $G_{00}=3=H_{00},$ $%
G_{01}=0=H_{10},$ $G_{10}=4=H_{01},$ and $G_{11}=2=H_{11}$ in matrices (\ref%
{AandBmatrices}). From (\ref{deltas}) we find $\Delta _{1}=1,$ $\Delta
_{2}=2 $ and $\Delta _{3}=1$ and we obtain $\gamma \leq \cos ^{-1}(1/3)$ for
a transition to the classical NE. Thus, for this PD game, to generate a
non-classical NE the entanglement parameter $\gamma $ should be greater than 
$\cos ^{-1}(1/3)$.

\subsubsection*{Stag Hunt}

The game of Stag Hunt (SH) \cite{Rasmusen} is encountered in the problems of
social cooperation. For example, if two hunters are hunting for food, in a
situation where they have two choices, either to hunt together and kill a
stag, which provides a large meal, or become distracted and hunt rabbits
separately instead, which while tasty, make a substantially smaller meal.
Hunting a stag of course is quite challenging and the hunters need to
cooperate with each other in order to be successful. The game of SH has
three classical NE, two of which are pure and one is mixed. The two pure NE
correspond to the situation where both hunters hunt the stag as a team or
where each hunts rabbits by himself.

The SH game can be defined by the conditions $\Delta _{3}>\Delta _{2}>0$ and 
$\Delta _{1}+\Delta _{2}>0$ and $\Delta _{3}>\Delta _{1}+\Delta _{2}$. In
the classical (mixed-strategy) version of this game three NE (two pure and
one mixed) appear consisting of $(x^{\ast },y^{\ast })=(0,0)$, $(x^{\ast
},y^{\ast })=(1,1)$ and $(x^{\ast },y^{\ast })=(\frac{\Delta _{2}}{\Delta
_{3}},\frac{\Delta _{2}}{\Delta _{3}})$. From Eq.~(\ref%
{eq:NEEPREmbeddedStagHunt}) and the defining conditions of SH game we notice
that both the strategy pairs $(0,0)$ and $(1,1)$ also remain NE in the
quantum game for an arbitrary $\gamma $. Eq.~(\ref%
{eq:AlicePayoffQuantumGameClassical}) give the players' payoffs at these NE
as follows:

\begin{eqnarray}
\Pi _{A}(0,0) &=&\frac{1}{2}\left[ G_{00}+G_{11}-\cos \gamma (G_{00}-G_{11})%
\right] =\Pi _{B}(0,0), \\
\Pi _{A}(1,1) &=&\frac{1}{2}\left[ G_{00}+G_{11}+\cos \gamma (G_{00}-G_{11})%
\right] =\Pi _{B}(1,1),
\end{eqnarray}%
which assume the values $G_{11}$ and $G_{00}$ at $\gamma =0$, respectively.
When $\gamma =\frac{\pi }{2}$ we have $\Pi _{A}(0,0)=\Pi _{A}(1,1)=\frac{1}{2%
}(G_{00}+G_{11})=\Pi _{B}(1,1)=\Pi _{B}(0,0)$. For the mixed NE for the
quantum SH game we require from Eq.~(\ref{eq:NEEPREmbeddedStagHunt}), $%
\Delta _{3}\{2y^{\ast }-1\}-\cos \gamma (\Delta _{1}+\Delta _{2})=0$ or 
\begin{equation}
x^{\ast }=\frac{\cos \gamma (\Delta _{1}+\Delta _{2})+\Delta _{2}-\Delta _{1}%
}{2\Delta _{3}}=y^{\ast },
\end{equation}%
which returns the classical mixed NE of $(\frac{\Delta _{2}}{\Delta _{3}},%
\frac{\Delta _{2}}{\Delta _{3}})$ at zero entanglement. Depending on the
amount of entanglement, the pair $(x^{\ast },y^{\ast })$, however, will
shift themselves between $\frac{\Delta _{2}}{\Delta _{3}}$ and $\frac{\Delta
_{2}-\Delta _{1}}{2\Delta _{3}}$. Players' payoffs at this shifted NE can be
obtained from Eq.~(\ref{eq:AlicePayoffQuantumGameClassical}). Consider a
particular example of SH by taking $G_{00}=10=H_{00},$ $G_{01}=0=H_{10},$ $%
G_{10}=8=H_{01},$ and $G_{11}=7=H_{11}$ in matrices (\ref{AandBmatrices}).
From (\ref{deltas}) we find $\Delta _{1}=-2,$ $\Delta _{2}=7$ and $\Delta
_{3}=9$. At $\gamma =\frac{\pi }{2}$ we have $\Pi _{A}(0,0)=\Pi _{A}(1,1)=%
\frac{17}{2}=\Pi _{B}(1,1)=\Pi _{B}(0,0)$. That is, the players' payoffs at
the NE strategy pair $(0,0)$ are increased from $7$ to $\frac{17}{2}$ while
at the NE strategy pair $(1,1)$ these are decreased from $10$ to $\frac{17}{2%
}$. The mixed NE in the classical game is at $x^{\ast }=\frac{7}{9}=y^{\ast
} $ whereas it shifts to $\frac{1}{2}$ at $\gamma =\frac{\pi }{2}$.

\section*{Discussion}

The EPR type setting for playing a quantum version of a two-player
two-strategy game is explored using the formalism of Clifford geometric
algebra (GA), used for the representation of the quantum states, and the
calculation of observables. We find that analyzing quantum games using GA
comes with some clear benefits, for instance, improved perception of the
quantum mechanical situation involved and particularly an improved
geometrical visualization of quantum operations. To obtain equivalent
results using the familiar algebra with Pauli matrices would be possible but
obscures intuition. We also find that an improved geometrical visualization
becomes helpful in significantly simplifying quantum calculations.

We find that by using an EPR type setting we \ produce a faithful embedding
of symmetric mixed-strategy versions of classical two-player two-strategy
games into its quantum version, and that GA provides a simplified formalism
over the field of reals for describing quantum states and measurements.

For a general two-player two-strategy game, we find the governing equation
for a strategy pair forming a NE and the associated payoff relations. We
find that at zero entanglement the quantum game returns the same pair(s) of
NE as the classical mixed-strategy game, while the payoff relations in the
quantum game reduce themselves to their bilinear form corresponding to a
mixed-strategy classical game. We find that, within our GA based analysis,
even though the requirement to properly embed a classical game puts
constraints on the possible quantum states allowing this, we still have a
degree of freedom, available with the entanglement angle $\gamma $, with
which we can generate new NE. As a specific example the PD was found to have
a NE of $(x^{\ast },y^{\ast })=(1,1)$ at high entanglement.

Analysis of quantum PD game in this paper can be compared with the results
developed for this game in Ref.~\cite{IqbalCheon} also using an EPR type
setting, directly from a set of non-factorizable joint probabilities.
Although Ref.~\cite{IqbalCheon} and the present paper both use an EPR type
setting, they use non-factorizability and entanglement for obtaining a
quantum game, respectively. Our recent work \cite{NFJP2010} has observed
that Ref.~\cite{IqbalCheon} does not take into consideration a symmetry
constraint on joint probabilities that is relevant both when joint
probabilities are factorizable or non-factorizable. When this symmetry
constraint is taken into consideration, an analysis of quantum PD game
played using an EPR setting does generate a non-classical NE in agreement
with the results in this paper.



\section*{Acknowledgments}

\bibliographystyle{plain}
\bibliography{quantum}





\end{document}